\documentclass[final,unsortedaddress,twocolumn,floatfix,%
  prb,amsmath,amssymb,showpacs]{revtex4}

\usepackage{latexsym}
\usepackage{epsfig}
\usepackage{natbib}
\usepackage{amssymb}
\usepackage{amsmath}
\usepackage{braket}
\usepackage{multirow}

\begin{document}

\title{
  Adiabatic and Non--Adiabatic Contributions to the Free Energy from the
  Electron--Phonon Interaction for Na, K, Al, and Pb
}

\date{\today}

\author{N.~Bock}
\email{nbock@lanl.gov}

\affiliation{
  Theoretical Division, Los Alamos National Laboratory, Los Alamos, New Mexico
  87545
}

\author{D.~Coffey}

\affiliation{
  Dept. of Physics, Buffalo State College, Buffalo, New York 14222
}

\author{Duane~C.~Wallace}

\affiliation{
  Theoretical Division, Los Alamos National Laboratory, Los Alamos, New Mexico
  87545
}



\begin{abstract}

We calculate the adiabatic contributions to the free energy due to the
electron--phonon interaction at intermediate temperatures, $0 \leqslant k_{B} T
< \epsilon_{F}$ for the elemental metals Na, K, Al, and Pb. Using our previously
published results for the nonadiabatic contributions we show that the adiabatic
contribution, which is proportional to $T^{2}$ at low temperatures and goes as
$T^{3}$ at high temperatures, dominates the nonadiabatic contribution for
temperatures above a cross--over temperature, $T_{c}$, which is between 0.5 and
0.8 $T_{m}$, where $T_{m}$ is the melting temperature of the metal. The
nonadiabatic contribution falls as $T^{-1}$ for temperatures roughly above the
average phonon frequency.

\vspace{0.2cm}

\noindent
LA-UR-05-6476

\end{abstract}

\maketitle

\section{
  Introduction
}

The crystal free energy consists of contributions from the static lattice
potential, the phonons, the electronic excitations, phonon--phonon interactions,
and electron-phonon interactions (\citet{Wallace_02:SPCL}, eq. (20.1)). The last
term is generally the smallest, but little is known about the size of this
contributions except at very low temperatures. We therefore study the
electron--phonon free energy, $F_{ep}$, in order to assess its contribution to
the thermodynamic properties of metals at all temperatures to melting.

The total electron--phonon free energy was derived by \citet{Eliashberg_60_JETP}
in a paper on superconductivity. In this formulation, the electron--phonon
interaction in the crystal ground state is double counted. The ground state
subtraction, which corrects for the ground state double counting, was derived in
\citet{Wallace_72:TOC}, eq. (25.26). It is well known that electron--phonon
interactions contribute to the low temperature electronic specific heat
$C_{el}$, according to

\begin{equation}
  \label{eq_specific_heat}
  C_{el} = \Gamma_{bs} \left( 1 + \lambda \right) T.
\end{equation}

\noindent
The bare electron contribution is $\Gamma_{bs} T$, where $\Gamma_{bs}$ is
proportional to the electronic density of states at the Fermi level determined
by bandstructure. The nonadiabatic electron--phonon contribution gives the term
in $\lambda$, and can be quite large, e.g. $\lambda \approx 1.5$ for lead
\cite{Allen_87,Bock_05}. In the early years, many theoretical calculations of
$\lambda$ were carried out for the simple metals, showing rather good agreement
between theory and experiment.  These results were reviewed by
\citet{Grimvall_76}, Tables III-VI; see also \citet{Wallace_72:TOC}, Tables 27
and 28. Grimvall \cite{Grimvall_76} evaluated the Eliashberg formula for lead,
and concluded that $F_{ep}$ vanishes at temperatures above the phonon
characteristic temperature. This conclusion will be revised in the present
report.

The adiabatic approximation \cite{Born_Oppenheimer_27} rests on the expansion of
the coupled nuclear--electron Hamiltonian in powers of $m/M$, the ratio of
electron to nuclear mass. When terms of order $m/M$ are neglected, the
electronic wavefunctions do not see the nuclear motion, and depend only on the
static nuclear positions \cite{Born_Oppenheimer_27}. The case of lattice
dynamics is treated by \citet{Born_56:DTCL}, p. 171. In the partition function,
the same property is revealed by treating the operation of the nuclear kinetic
energy on the electronic wavefunctions as a small effect \cite{Zwanzig_57}. When
this operation is neglected entirely, which is appropriate at high temperatures
where the nuclear motion is classical, only the adiabatic part $F^{ad}$ survives
in $F_{ep}$ (\citet{Wallace_02:SPCL}, p. 91--93). It is therefore seen that the
electronic states are not mixed in the adiabatic part, and also that $F^{ad}$ is
the dominant part at high temperatures. The nonadiabatic part $F^{na}_{1} +
F^{na}_{2}$ arises from mixing of electronic states by the nuclear kinetic
energy, and becomes unimportant at high temperatures.

\citet{Allen_Heine_76} separated electron--phonon effects into adiabatic and
nonadiabatic contributions, and studied the adiabatic part of the electron
energy shifts. For the Eliashberg formulation, \citet{Allen_Hui_80} studied the
adiabatic contribution to the high temperature specific heat. Present results
will be compared with Allen and Hui in Sec. \ref{sec_results}. The leading
correction to the linear temperature dependence in the specific heat given in
eq. (\ref{eq_specific_heat}) is a nonadiabatic term which goes as $T^{3} \ln T$.
This has been studied extensively in the literature starting with
\citet{Buckingham_51} and \citet{Buckingham_Schafroth_54}. See reference
\cite{Danino_Overhauser_82} and \cite{Coffey_Pethick_88} and references therein
for more recent work.

We have previously addressed the problem of calculating the contribution to the
free energy from the electron--phonon interaction \cite{Bock_05}. We found the
nonadiabatic part of this contribution, $F_{2}^{na}$, to second order in the
interaction. We calculated $F_{2}^{na}$ for four nearly--free--electron metals,
Na, K, Al, and Pb, for temperatures between zero and roughly 1.5 times the
melting temperature.

As in the calculation of the nonadiabatic contribution, our calculations are
done for a constant density $\rho$ to eliminate concern for the density
dependence of phonon frequencies and electron--phonon interaction matrix
elements. The density is that at the temperature $T_{\rho}$, where the phonon
frequencies are measured. The melting temperature at this density is higher than
the customary zero--pressure melting temperature. Our calculations cover the
range from $T = 0$ to above $T_{m}$.

The paper is organized as follows. In Sec. \ref{sec_ep_free_energy} we present
the expression for the adiabatic and nonadiabatic parts of $F_{ep}$, and we
discuss the formulation for metals in general, the simplification for the
nearly--free--electron metals studied here, and the groundstate subtraction.
Asymptotic temperature dependences are derived for the adiabatic part. In Sec.
\ref{sec_results} our results are presented with a detailed discussion of the
numerical methods used, the adiabatic and nonadiabatic parts are compared, and
our results are compared with previous work. Our conclusions are summarized in
\ref{sec_conclusions}.

\section{
  Adiabatic contribution to the Electron--Phonon Free Energy
}
\label{sec_ep_free_energy}

\subsection{
  Analytic Form of the Free Energy
}

The electron--phonon contribution to the free energy can be written in three
pieces, $F_{ep} = F^{ad} + F_{1}^{na} + F_{2}^{na}$,

\begin{widetext}
\begin{eqnarray}
  \label{eq_F_ad}
  \frac{F^{ad}}{N} & = & \sum_{\vec{p} \vec{k} \vec{Q} \lambda}
  \frac{\hbar^{2}}{N^{2} M}
  \frac{n_{\vec{k} \lambda} + \frac{1}{2}}{\hbar \omega_{\vec{k} \lambda}}
  \left( f_{\vec{p}} - g_{\vec{p}} \right)
  \left\{
    \frac{\left[ \left( \vec{k} + \vec{Q} \right) \cdot
      \hat{\eta}_{\vec{k} \lambda} \right]^{2}
      \left[ U \left( \vec{k} + \vec{Q} \right) \right]^{2} }
      {\epsilon_{\vec{p}} - \epsilon_{\vec{p} + \vec{k} + \vec{Q}}}
    -
    \frac{\left[ \vec{Q} \cdot \hat{\eta}_{\vec{k} \lambda} \right]^{2}
      \left[ U \left( \vec{Q} \right) \right]^{2} }
      {\epsilon_{\vec{p}} - \epsilon_{\vec{p} + \vec{Q}}}
  \right\} \\
  \label{eq_F_1_na}
  \frac{F_{1}^{na}}{N} & = & \sum_{\vec{p} \vec{k} \vec{Q} \lambda}
  \frac{\hbar^{2}}{N^{2} M}
  \hbar \omega_{\vec{k} \lambda}
  \left( n_{\vec{k} \lambda} + \frac{1}{2} \right)
  \frac{f_{\vec{p}}}{\epsilon_{\vec{p}} - \epsilon_{\vec{p} + \vec{k} + \vec{Q}}}
  \frac{\left[ \left( \vec{k} + \vec{Q} \right) \cdot
    \hat{\eta}_{\vec{k} \lambda} \right]^{2}
    \left[ U \left( \vec{k} + \vec{Q} \right) \right]^{2}}
    {\left[ \epsilon_{\vec{p}} - \epsilon_{\vec{p} + \vec{k} + \vec{Q}} \right]^{2}
     - \left[ \hbar \omega_{\vec{k} \lambda} \right]^{2}} \\
  \label{eq_F_2_na}
  \frac{F_{2}^{na}}{N} & = & \sum_{\vec{p} \vec{k} \vec{Q} \lambda}
  \frac{\hbar^{2}}{2 N^{2} M}
  f_{\vec{p}} \left( 1 - f_{\vec{p} + \vec{k} + \vec{Q}} \right)
  \frac{\left[ \left( \vec{k} + \vec{Q} \right) \cdot
    \hat{\eta}_{\vec{k} \lambda} \right]^{2}
    \left[ U \left( \vec{k} + \vec{Q} \right) \right]^{2}}
    {\left[ \epsilon_{\vec{p}} - \epsilon_{\vec{p} + \vec{k} + \vec{Q}} \right]^{2}
     - \left[ \hbar \omega_{\vec{k} \lambda} \right]^{2}}.
\end{eqnarray}
\end{widetext}

\noindent
Our results are calculated and quoted per atom. Throughout this paper we will
use the following nomenclature: $f_{\vec{p}}$ is the Fermi--Dirac distribution
function at finite temperature and $g_{\vec{p}}$ is the same at $T = 0$.
$n_{\vec{k} \lambda}$ is the Bose--Einstein distribution function at finite
temperature and $\hat{\eta}_{\vec{k} \lambda}$ is the polarization vector of the
phonon branch $\lambda$ for wave vector $\vec{k}$ which is inside the Brillouin
zone.  $\vec{Q}$ is a reciprocal lattice vector and $\omega_{\vec{k} \lambda}$
is the frequency of a phonon mode. $U (\vec{k} + \vec{Q})$ is the Fourier
transform of the pseudopotential for momentum transfer $\vec{k} + \vec{Q}$

To remind the reader of the physical meaning of the three terms in eqs.
(\ref{eq_F_ad} - \ref{eq_F_2_na}) (a more detailed discussion can be found in
\citet{Bock_05}) a quick summary: Eq. (\ref{eq_F_ad}), $F^{ad}$, expresses the
thermally averaged vibrational contributions to the excited electronic energies.
Eqs.  (\ref{eq_F_1_na}) and (\ref{eq_F_2_na}), $F_{1,2}^{na}$, describe the
non--adiabatic corrections to all electronic energy levels and take into account
the mixing of electron states due to the ion motion.

In the free energy formulation, the electron--phonon interaction is treated in
second order perturbation theory \cite{Wallace_02:SPCL, Eliashberg_60_JETP,
Wallace_72:TOC, Allen_87, Bock_05, Grimvall_76, Allen_Heine_76, Allen_Hui_80,
Danino_Overhauser_82, Coffey_Pethick_88}. For a general metal, the electrons are
presumed to have band structure, and the energy denominators are band electron
energies $E_{\vec{k}}$. For nearly--free--electron metals, band structure
effects may be treated in pseudopotential perturbation theory, where in zeroth
order the electron energies are the free electron energies $\epsilon_{\vec{k}}$.
This is why free electron energies appear in the denominator of eqs.
(\ref{eq_F_ad})--(\ref{eq_F_2_na}). In these equations, the band structure
effects, to second order in the pseudopotential, are contained in the last term
in brackets in eq. (\ref{eq_F_ad}). Pseudopotential perturbation theory has been
extensively developed over many years, and pseudopotential parameters have been
calibrated to experimental data such as equilibrium density and bulk modulus. We
use these calibrated pseudopotentials here, so that our models have no free
parameters.

Finally we need to clarify the problem of groundstate double counting. The
potential for the nuclear motion, the ``adiabatic potential'', is precisely the
electronic groundstate energy as a function of static nuclear positions. Since
this potential has been put into the phonon Hamiltonian, the electronic
groundstate energy has to be subtracted from the electron--phonon Hamiltonian.
In this way the electronic statistical mechanics expresses only electronic
excitations from the groundstate. The groundstate subtraction is an adiabatic
effect, and is expressed by the term $\left( - g_{\vec{p}} \right)$ in eq.
(\ref{eq_F_ad}).  The decomposition of the total Hamiltonian for a metal
crystal, and derivation of the corresponding free energy, including eqs.
(\ref{eq_F_ad})--(\ref{eq_F_2_na}), is given in \citet{Wallace_02:SPCL}.

\subsection{
  Analytic temperature dependence
}

The temperature dependence of $F^{ad}$, eq. (\ref{eq_F_ad}), arises from the
product of the Fermi--Dirac factor $\left( f_{\vec{p}} - g_{\vec{p}} \right)$
and the phonon factor $\left( n_{\vec{k} \lambda} + 1/2 \right)$. The value at
$T = 0$, i.e. the constant term in $F^{ad}$, vanishes because $\left(
f_{\vec{p}} - g_{\vec{p}} \right)$ vanishes at $T = 0$. Using a Sommerfeld
expansion (e.g. pp. 45 in \citet{Ashcroft_76:SSP}) of eq. (\ref{eq_F_ad}), the
Fermi--Dirac factor gives a quadratic temperature dependence in leading order,
plus higher order terms which are of relative order $\left( k_{B} T /
\epsilon_{F} \right)^{2}$ and so can be neglected at temperatures to well above
$T_{m}$. The phonon factor reduces to the constant $1/2$ at very low
temperatures, i.e. at $k_{B} T \ll$ the mean phonon energy, $\left< \hbar \omega
\right>$, so that the net dependence at very low temperatures is

\begin{equation}
  \label{eq_low_T_T2}
  F^{ad} = B_{2} \left( k_{B} T \right)^{2}.
\end{equation}

\noindent
At higher temperatures, i.e. at $k_{B} T \gtrsim \left< \hbar \omega \right>$,
each phonon factor can be expanded as

\begin{equation}
  \label{eq_n_plus_half}
  \left( n_{\vec{k}} + \frac{1}{2} \right)
  =
  \frac{1}{\beta \hbar \omega}
    + \frac{\beta \hbar \omega}{12}
    + \frac{\left( \beta \hbar \omega \right)^{3}}{720}
    + \cdots
\end{equation}

\noindent
Hence the temperature dependence at $k_{B} T \gtrsim \left< \hbar \omega
\right>$ becomes

\begin{equation}
  \label{eq_low_T_T3}
  F^{ad} = B_{3} \left( k_{B} T \right)^{3} + \cdots
\end{equation}

\noindent
As we will show in a later section, our numerical results confirm this analytic
temperature dependence.

\section{
  Results
}
\label{sec_results}

\subsection{
  Phonon Model
}
\label{sec_results_phonon_model}

\begin{figure}
  \psfig{file=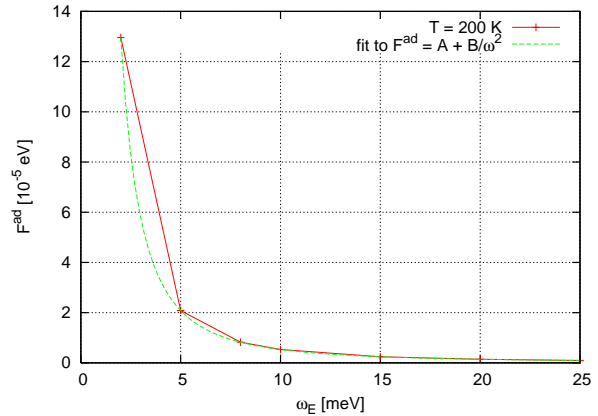,angle=-90,width=8cm}
  \caption{
    \label{fig_set_01}
    (Color online) $\omega_{E}$--dependence of $F^{ad}$ for Na at $T$ = 200 K.
  }
\end{figure}
%
%
%

When we calculated the non--adiabatic contributions we found that we could
replace the phonon dispersion with a constant frequency with high accuracy which
simplified our numerical evaluation. In fact, we found that the result of our
calculation was not sensitive to the value of this frequency, $\omega_{E}$, and
that an Einstein model gives the correct temperature dependence. We reasoned
that this insensitivity is due to the fact that the integrands of $F_{1,2}^{na}$
are finite and well--behaved as a function of $\omega_{\vec{k} \lambda}$. The
adiabatic contribution on the other hand diverges as $\omega_{E} \rightarrow 0$
as illustrated in Fig.  \ref{fig_set_01} and is therefore very sensitive to the
choice of $\omega_{E}$. Although the Einstein model will give the correct
temperature dependence at low and high temperatures as listed in eqs.
(\ref{eq_low_T_T2}) and (\ref{eq_low_T_T3}) we have no way to find the Einstein
frequency which will give us the correct magnitude without doing a calculation
with the full phonon dispersion. This is a serious problem as can be seen from
the extreme dependence on $\omega_{E}$ as shown in Fig. \ref{fig_set_01}.  Since
in the case of $F^{ad}$ the calculation using the full phonon dispersion is not
prohibitive, we chose for this report to use a fully $\vec{k}$--dependent phonon
dispersion.


In the following we calculated the phonon dispersion and eigenvectors from
Born--von K\'{a}rm\'{a}n force constants to get a more realistic representation
of the real phonon spectrum with its three branches (the force constants were
taken from \citet{Landolt_81}).


\subsection{
  Numerical Techniques
}

Eq. (\ref{eq_F_ad}) is written in terms of sums over the electron and phonon
momenta. We are using a free electron dispersion spectrum which is isotropic in
the electron momentum. This combined with the fact that $\left| \vec{p} \right|$
is bounded by $\left( f_{\vec{p}} - g_{\vec{p}} \right)$ from above and below to
$\left| \vec{p} \right| \approx p_{F}$ to within a few $k_{B} T$ makes it
numerically more convenient for us to rewrite $\sum_{\vec{p}}$ as an integral,

\begin{equation}
\sum_{\vec{p}} = \frac{N V_{A}}{\left( 2 \pi \right)^{3}} \int d^{3} p.
\end{equation}

\noindent
The sum over phonon momenta however is evaluated more conveniently as a sum
because of the two terms in the curly brackets in $F^{ad}$. The momentum
transfer $\vec{k} + \vec{Q}$ enters the second term only with the reciprocal
lattice vector $\vec{Q}$. Were we to integrate over the final electron momentum,
$\vec{p'} = \vec{p} + \vec{k} + \vec{Q}$, the second term would introduce
discrete steps into the integrand as we change $\vec{p'}$. This is very
difficult to handle numerically. Summing over $\vec{k}$ and $\vec{Q}$ instead
converges much faster. The expression we evaluated is given by

\begin{widetext}
\begin{equation}
  \label{eq_F_ad_einstein}
  \frac{F^{ad}}{N} =
  \frac{ V_{A} }{ \left( 2 \pi \right)^{3} }
  \,\,
  \mathcal{P}
  \int d\vec{p}
  \,\,
  \sum_{\vec{k} \vec{Q} \lambda}
  \frac{\hbar^{2}}{N M}
  \frac{n_{\vec{k} \lambda} + \frac{1}{2}}{\hbar \omega_{\vec{k} \lambda}}
  \left( f_{\vec{p}} - g_{\vec{p}} \right)
  \left\{
    \frac{\left[ \left( \vec{k} + \vec{Q} \right) \cdot
      \hat{\eta}_{\vec{k} \lambda} \right]^{2}
      \left[ U \left( \vec{k} + \vec{Q} \right) \right]^{2} }
      {\epsilon_{\vec{p}} - \epsilon_{\vec{p} + \vec{k} + \vec{Q}}}
    -
    \frac{\left[ \vec{Q} \cdot \hat{\eta}_{\vec{k} \lambda} \right]^{2}
      \left[ U \left( \vec{Q} \right) \right]^{2} }
      {\epsilon_{\vec{p}} - \epsilon_{\vec{p} + \vec{Q}}}
  \right\} \\
\end{equation}
\end{widetext}

\noindent
where $\mathcal{P}$ denotes the Cauchy principal value.

\begin{figure}
  \psfig{file=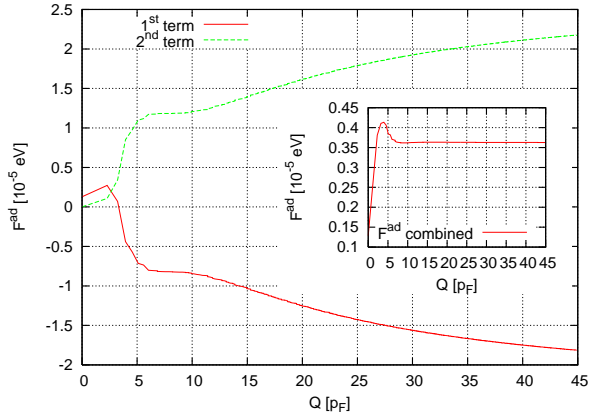,angle=-90,width=8cm}
  \caption{
    \label{fig_set_15}
    (Color online) Convergence behavior of $F^{ad}$ for Na at $T$ = 200 K. The
    inset shows the combined terms.
  }
\end{figure}

At this point we would like to address the question of convergence in our
calculation since the sum on $\vec{k} + \vec{Q}$ is not obviously bounded by an
upper limit. We had found for the two nonadiabatic parts strong convergence as a
function of $\vec{p'}$ due to a combination of pseudopotential factor and the
energy denominator. In the case of the adiabatic contribution however, this is
not quite so obvious. The energy denominator is only linear in the energy
difference and we therefore expect the convergence of the two single terms in
$F^{ad}$ to go as $Q^{-2}$, which is quite weak. We do suspect however that the
two terms in the curly brackets will exhibit cancellation to some degree and
might improve the convergence behavior of $F^{ad}$. In Fig.  \ref{fig_set_15} we
plot the two terms separately and combined as a function of magnitude of
$\vec{Q}$. Despite the slow convergence behavior of the terms by themselves,
their combination converges strongly.

\subsection{
  Comparing the adiabatic with the nonadiabatic results
}

\begin{figure}
  \psfig{file=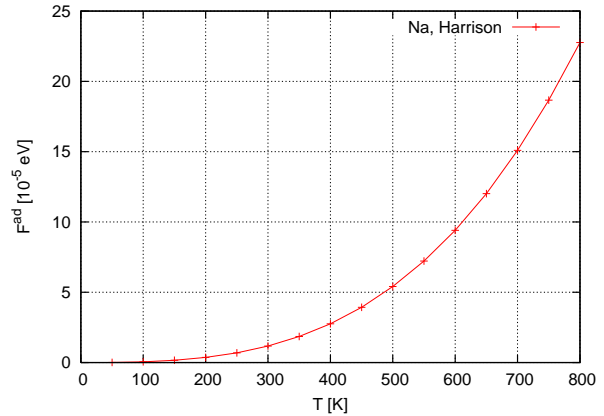,angle=-90,width=8cm}
  \caption{
    \label{fig_set_07}
    $F^{ad}$ for Na using full phonons.
  }
\end{figure}

As mentioned earlier, we calculated $F^{ad}$ using force constant models for the
phonon dispersion for the elemental metals Na, K, Al, and Pb. Our result for Na
is shown in Fig.  \ref{fig_set_07}. A polynomial fit of the temperature
dependence in the low-- and high--temperature regime shows that our initial
analysis was correct and $F^{ad} \propto T^{2}$ for low temperatures and that it
turns into $\propto T^{3}$ for higher temperatures.

The sign of $F^{ad}$ for all metals studied here except Al is positive. It is
not obvious from eq. (\ref{eq_F_ad_einstein}) why this is the case. We can
speculate that the sign is determined by the relative magnitudes of the two
terms in parentheses in $F^{ad}$ since we know from our convergence study (Fig.
\ref{fig_set_15}) that the two terms are roughly of equal strength but of
opposite sign. The difference between the two terms is the scattering momentum
which is the full phonon momentum in the first term ($\vec{k} + \vec{Q}$) and
the reciprocal lattice vector only in the second term ($\vec{Q}$).

The Fermi--Dirac factor $\left( f_{\vec{p}} - g_{\vec{p}} \right)$ ensures that
the first electron momentum, $\vec{p}$, is essentially confined to the Fermi
surface, $p_{F}$. For the denominators to be small, the second electron
momentum, ($\vec{p} + \vec{k} + \vec{Q}$) and ($\vec{p} + \vec{Q}$), has to
therefore lie close to the Fermi surface. Although the exact shape of the Fermi
surface can be quite complicated we would like to restrict this argument to a
spherical Fermi surface since we are using the pseudopotential formalism in this
calculation and do not take separate bands into account. In this case it is
plausible to speculate that a change of the radius of the Fermi surface might be
able to effect an overall sign change in $F^{ad}$.

\begin{figure}
  \psfig{file=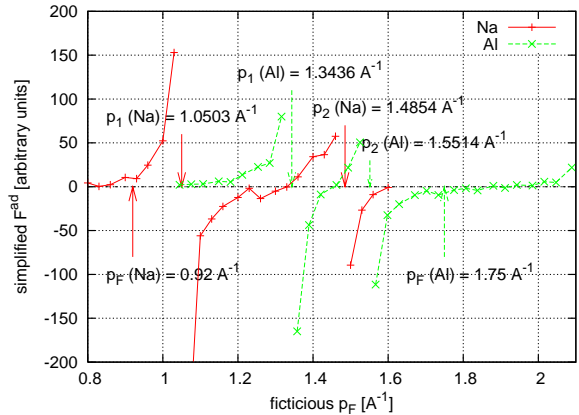,angle=-90,width=8cm}
  \caption{
    \label{fig_set_16}
    (Color online) Dependence of sign of $F^{ad}$ as a function of $p_{F}$ for
    Na at $T$ = 200 K and for Al at $T$ = 500 K. For illustration purposes, the
    lengths $p_{1,2} = 1/2 \left| \vec{Q}_{1,2} \right|$ of the first two
    reciprocal lattice vectors are shown.
  }
\end{figure}

We explore this possibility by means of a ``simplified $F^{ad}$'', given by eq.
(\ref{eq_F_ad_einstein}) with the pseudopotential set to a constant and with a
``fictitious $p_{F}$'' that is allowed to vary. This amounts to pretending that
we can change the number of free electrons, hence change the size of the Fermi
surface, without changing anything else. In this case the essential part of the
second term in eq. (\ref{eq_F_ad_einstein}) is the $\vec{p}$--angle average of
$(\epsilon_{\vec{p}} - \epsilon_{\vec{p} + \vec{Q}} )^{-1}$ with $\left|
\vec{p} \right|$ equal to the fictitious $p_{F}$ and with $Q$ any reciprocal
lattice vector. This integral diverges as the fictitious $p_{F}$ passes through
$p_{n} = 1/2 \,\, Q_{n}$, where $Q_{1}$ is the magnitude of the first reciprocal
lattice vector, and so on for $Q_{2}$, etc. At the same time, the first term in
eq. (\ref{eq_F_ad_einstein}) is well behaved as a function of the fictitious
$p_{F}$. Numerical evaluation of the simplified $F^{ad}$ is shown in Fig.
\ref{fig_set_16} for Na and Al, and reveals the expected logarithmic
discontinuities at $p_{n} = 1/2 \,\, Q_{n}$. Fig \ref{fig_set_16} supports the
view that the sign of $F^{ad}$ depends directly on the size of the Fermi surface
relative to the Brillouin zone surface.

\begin{figure}
  \psfig{file=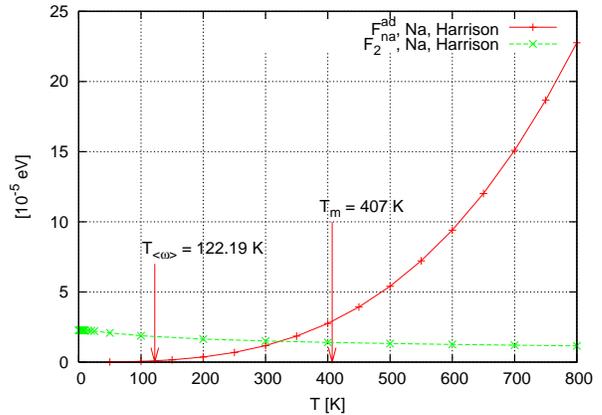,angle=-90,width=8cm}
  \caption{
    \label{fig_set_17}
    (Color online) $F^{ad}$ and $F_{2}^{na}$ for Na. The cross--over temperature
    is shown.
  }
\end{figure}

Our results for Na for $F^{ad}$ including our previous results for $F_{2}^{na}$
are shown in Fig. \ref{fig_set_17}. At low temperatures, the adiabatic
contribution vanishes and the nonadiabatic contribution approaches a constant.
In the low temperature regime, the free energy is dominated by the nonadiabatic
contribution. In the high temperature regime, the adiabatic contribution
increases as $T^{3}$ and the nonadiabatic contribution slowly vanishes. The
adiabatic contribution dominates in this temperature regime. At about $\approx
320$ K there is a cross--over between the two contributions. For reference we
included the temperature of an average phonon frequency and the melting
temperature of Na in the graph. The cross--over temperature is well between
either one of these two temperatures.

\begin{figure}
  \psfig{file=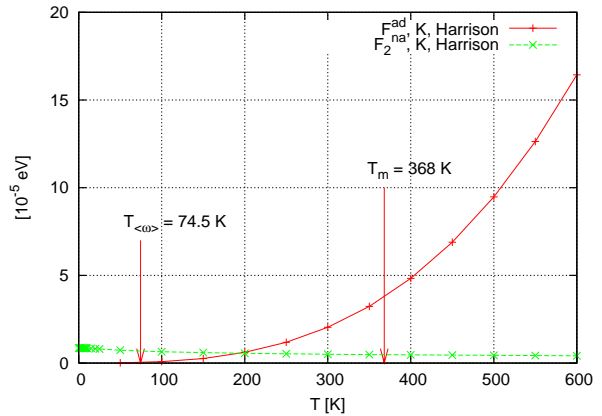,angle=-90,width=8cm}
  \caption{
    \label{fig_set_18}
    (Color online) $F^{ad}$ and $F_{2}^{na}$ for K. The cross--over temperature
    is shown.
  }
\end{figure}

\begin{figure}
  \psfig{file=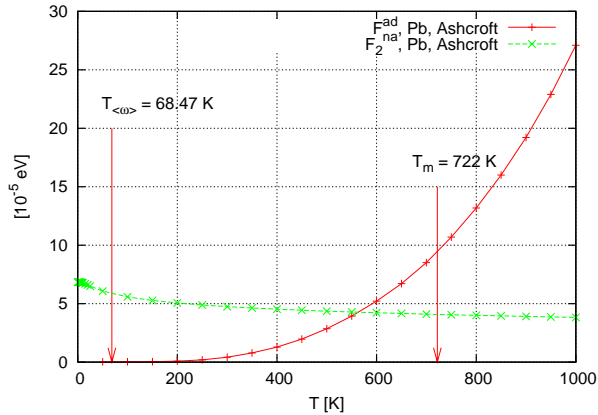,angle=-90,width=8cm}
  \caption{
    \label{fig_set_20}
    (Color online) $F^{ad}$ and $F_{2}^{na}$ for Pb The cross--over temperature
    is shown.
  }
\end{figure}

Fig. \ref{fig_set_18} shows our results for K. We find the same qualitative
behavior. The cross--over temperature is always well between the two reference
temperatures. The same is seen for our results for Pb, shown in Fig.
\ref{fig_set_20}.

\begin{figure}
  \psfig{file=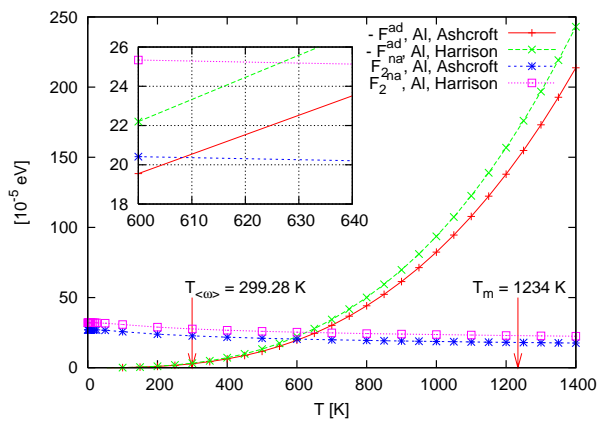,angle=-90,width=8cm}
  \caption{
    \label{fig_set_19}
    (Color online) $F^{ad}$ and $F_{2}^{na}$ for Al. Note that $F^{ad}$ is
    negative but drawn here as $-F^{ad}$. The cross--over temperature is shown.
    Both pseudopotentials are used. The inset shows the cross--over region. The
    datasets are labeled as in the larger plot.
  }
\end{figure}

In the case of Al we repeated the calculation for both pseudopotential models.
Our results are shown in Fig. \ref{fig_set_19}. The different pseudopotentials
effect a slightly different curvature of the temperature dependence in the
adiabatic contribution. As we found previously, the nonadiabatic contribution is
shifted in energy by about $5 \times 10^{-5}$ eV. The net result of these two
effects on the cross--over temperature (between the magnitudes of $F^{ad}$ and
$F^{na}_{2}$) is to shift this temperature by only roughly 20 K. The difference
between the two pseudopotential models is presumable well within the accuracy of
the pseudopotential method and our calculation itself. It does not make any
difference which model is chosen.

All of the cross--over temperatures are summarized in table
\ref{table_cross_temp}.

\begin{table}
\caption{
  \label{table_cross_temp}
  Cross--Over temperatures and other quantities
}
\begin{tabular}{lcccc}
\hline
\hline
  & 
    Na &
    K &
    Al &
    Pb \\
\hline
  structure &
    bcc &
    bcc &
    fcc &
    fcc \\
  $\left< \hbar \omega \right>$ [meV] &
    10.53 &
    6.42 &
    25.79 &
    5.90 \\
  $T_{melt}$ [K] &
    407 &
    368 &
    1234 &
    722 \\
  $T_{melt}^{P = 0}$ [K] &
    371.0 &
    336.4 &
    933.5 &
    600.6 \\
  $T_{c}$ [K] &
    325 &
    194 &
    609 (A), 627 (H) &
    563 \\
  $F^{ad} / F_{el} (T_{m})$ &
    -0.03 &
    -0.03 &
    0.19 (A), 0.22 (H) &
    -0.13 \\
  $S^{ad} / S_{el} (T_{m})$ &
    -0.045 &
    -0.045 &
    0.26 (A), 0.30 (H) &
    -0.20 \\
  $C^{ad} / C_{el} (T_{m})$ &
    -0.09 &
    -0.09 &
    0.45 (A), 0.50 (H) &
    -0.39 \\
\hline
\hline
\end{tabular}
\end{table}

In order to assess the overall importance of the adiabatic electron--phonon free
energy at high temperatures, we compare it with the bare electron free energy
$F_{el}$. The leading order Sommerfeld expansion of $F_{el}$, which is quite
accurate to $T_{m}$ for the nearly--free--electron metals, is

\begin{equation}
  F_{el} = -\frac{1}{2} \Gamma_{bs} T^{2}.
\end{equation}

\noindent
Since $F^{ad}$ increases approximately as $T^{3}$ at high $T$, the ratio $F^{ad}
/ F_{el}$ has its maximum magnitude at $T_{m}$. This ratio is listed in Table
\ref{table_cross_temp}, based on our calculations and the electronic density of
states of \citet{Moruzzi_Janak_Williams:CEPM}. Our corresponding values for
$S^{ad} / S_{el}$ at $T_{m}$, and $C^{ad} / C_{el}$ at $T_{m}$, are also listed
in Table \ref{table_cross_temp}. These results allow us to make estimates for
nearly--free--electron metals, of whether or not the adiabatic electron--phonon
contribution is significant in a given property of a given metal.

\subsection{
  Comparison with previous work
}

Previous work on the electron--phonon contribution was done on the entropy
instead of the free energy. The thermodynamic relation

\begin{equation}
S = - \frac{\partial F}{\partial T}
\end{equation}

\noindent
relates the two quantities. In order to see how our result for the free energy
affects the entropy, we took the derivative of the adiabatic and nonadiabatic
contributions and plotted the entropy for temperatures between 100 K and 1000 K
for lead in Fig. \ref{fig_set_30_b}. The nonadiabatic entropy slowly vanishes
with increasing temperature, whereas the adiabatic contribution increases. The
electron--phonon contribution to the entropy in lead was calculated by
\citet{Grimvall_EPI_81}. We found previously that for temperatures up to $T \leq
1.4 \,\, T_{E}$ the nonadiabatic contribution to the free energy is sufficient
to achieve good agreement with Grimvall's entropy calculation. We see now why.
The adiabatic contribution is too small to contribute noticeably to the total
entropy. At higher temperatures however, this will not be the case anymore and
the adiabatic contribution will become important. As we see from Fig.
\ref{fig_set_30_b}, $S^{ad}$ increases with temperature and at around $T = 240$
K $\left| S^{ad} \right| > \left| S_{2}^{na} \right|$. Grimvall's conclusion
that $S_{ep} (T \rightarrow  \infty) = 0$ is therefore not correct.  We also
notice that, maybe not surprisingly, the cross--over temperature for the free
energy is higher than the cross--over temperature for the entropy.

\begin{figure}
  \psfig{file=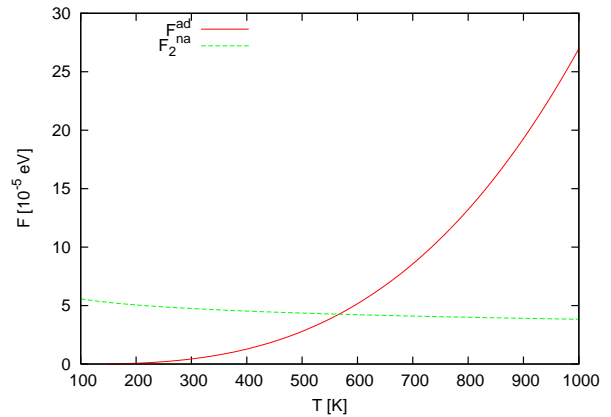,angle=-90,width=8cm}
  \caption{
    \label{fig_set_30_a}
    (Color online) The free energy contributions $F^{ad}$ and $F_{2}^{na}$ for
    Pb.
  }
\end{figure}

\begin{figure}
  \psfig{file=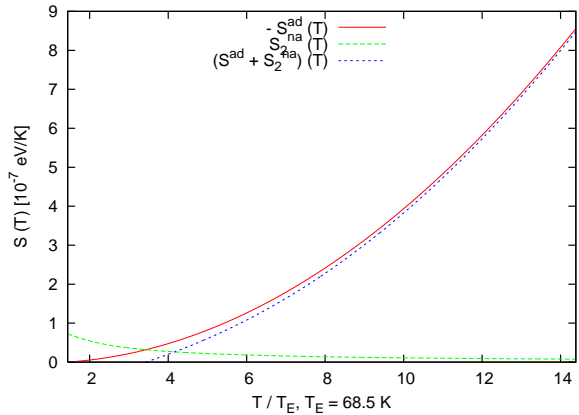,angle=-90,width=8cm}
  \caption{
    \label{fig_set_30_b}
    (Color online) The entropy contributions $S^{ad}$ and $S_{2}^{na}$ for Pb.
    The free energy is shown for comparison in Fig \ref{fig_set_30_a}.
  }
\end{figure}


\citet{Allen_Hui_80} studied the adiabatic electron--phonon specific heat as
derived from the Eliashberg free energy. They argued that for A15 metals, which
have narrow peaks in the electronic density of states, the adiabatic
contribution can be as large as the bare electronic specific heat at high
temperatures. Our results in Table \ref{table_cross_temp} are not inconsistent
with this possibility. Differences in our procedure and theirs prevent a further
comparison of results. First, their electron--phonon free energy does not
contain the groundstate subtraction. We note, however, that while the
groundstate subtraction contributes to the free energy and entropy at all
temperatures, its contribution to the specific heat vanishes at high
temperatures.  Second, our adiabatic specific heat is $C^{ad} = - T \left(
\partial^{2} F^{ad} / \partial T^{2} \right)_{V}$, while Allen and Hui remove
certain terms from this $C^{ad}$, to be placed in the bare phonon and bare
electron specific heats.  This, of course, is not an error but a matter of
choice. Nevertheless all the terms in question arise from eq. (\ref{eq_F_ad})
for $F^{ad}$ and are therefore genuine electron--phonon interaction terms.

\section{
  Conclusions
}
\label{sec_conclusions}

\subsection{
  Structure of the Free Energy
}

The exact free energy of a crystalline elemental metal can be expressed as
\cite{Wallace_02:SPCL,Bock_05}

\begin{eqnarray}
  \lefteqn{F = \Phi_{0} (V) + F_{el} (V, T) + F_{ep} (V, T)}
    \nonumber \\
  \label{eq_free_energy}
  & & \mbox{} + F_{ph} (V, T) + F_{anh} (V, T),
\end{eqnarray}

\noindent
where $\Phi_{0}$ is the static lattice potential, $F_{ph}$ represents
quasiharmonic phonons, $F_{anh}$ represents phonon--phonon interactions,
$F_{el}$ represents static--lattice electronic excitations, and $F_{ep}$ is the
remainder of the free energy expressed in terms of interactions between
electronic excitations and phonons. $F_{ep}$ is the smallest and most
complicated term in eq. (\ref{eq_free_energy}). In leading order thermodynamic
perturbation theory, $F_{ep}$ is composed of a nonadiabatic part, $F^{na} =
F_{1}^{na} + F_{2}^{na}$, and an adiabatic part, $F^{ad}$. The nonadiabatic part
arises from the mixing of electronic states due to the motion of the ions, and
strongly affects the lowest electronic excitations. Hence $F^{na}$ dominates
$F_{ep}$ at low temperatures. The adiabatic part, $F^{ad}$, expresses the
averaging over the thermal motion of the ions of each electronic excited state,
and is formally the leading contribution to $F_{ep}$ at high temperatures where
the ion motion is classical. Hence there exists a crossover temperature $T_{c}$,
below which the major contribution to $F_{ep}$ is $F^{na}$, and above which it
is $F^{ad}$.

\subsection{
  Sign and Magnitude Estimates
}

The following properties are characteristic of our results for Na, K, Al, and
Pb. We suppose these properties are common but not without exception among
metals in general.

\begin{enumerate}
  \item
    $F_{1}^{na}$ is negligible compared to $F_{2}^{na}$ for $0 \leq T \leq
    T_{m}$.
  \item
    $F_{2}^{na}$ is positive at all $T$. From a positive value at $T = 0$,
    $F_{2}^{na}$ decreases as $T^{2}$ at first, then decreases ever more slowly
    with increasing $T$ (Figs. 4, 6, 8, 9 of \citet{Bock_05}).
  \item
    At very low temperatures, $F_{2}^{na}$ is the only significant contribution
    to $F_{ep}$, and has the form $F_{2}^{na} = C_{2} + A_{2} T^{2}$, where
    $C_{2} > 0$ and $A_{2} < 0$. $C_{2}$ constitutes an entirely negligible
    contribution to the electronic groundstate energy. The temperature
    dependence is the same as $F_{el} = -(1/2) \Gamma_{bs} T^{2}$ at low
    temperatures, so that $F_{2}^{na}$ causes the well--known electron--phonon
    correction to the bare electronic specific heat. We have $C_{el} + C_{ep} =
    \left( \Gamma_{bs} - 2 A_{2} \right) T$, where $\left| 2 A_{2} / \Gamma_{bs}
    \right|$ ranges from a few percent to around 2.
  \item
    $F^{ad}$ can be of either sign. $F^{ad}$ is negligible compared to
    $F_{2}^{na}$ at low temperatures, but $\left| F^{ad} \right| > F_{2}^{na}$
    at $T > T_{c}$. This is because of the strong temperature dependence of
    $F^{ad} = B_{3} \left( k_{B} T \right)^{3}$ at $k_{B} T \gtrsim \left< \hbar
    \omega \right>$. Though $\left| F^{ad} / F_{2}^{na} \right|$ increases with
    temperature, $F_{2}^{na}$ it not always negligible at $T_{m}$ (Figs.
    \ref{fig_set_17} - \ref{fig_set_19}). On the other hand, $S_{2}^{na}$ is
    negligible compared to $\left| S^{ad} \right|$ at high temperatures (Fig.
    \ref{fig_set_30_b}).
  \item
    The ratio $\left| F^{ad} / F_{el} \right|$ reaches its maximum at $T_{m}$,
    where it ranges from a few percent to 0.2 (Table \ref{table_cross_temp}).
    The ratio $\left| S^{ad} / S_{el} \right|$ is larger, and $\left| C^{ad} /
    C_{el} \right|$ is larger still (Table \ref{table_cross_temp}), so that the
    adiabatic contribution to the specific heat can become important at high
    temperatures.
  \item
    The theory is a complex mixture of electronic excitations and phonons, hence
    the crossover temperature $T_{c}$ does not scale with either $\epsilon_{F}$
    or $\left< \hbar \omega \right>$. From our calculations, $k_{B} T_{c}$ is
    well above $\left< \hbar \omega \right>$, and $T_{c} / T_{m}$ is in the
    range 0.5 - 0.8.
  \item
    Based on our results shown in Fig. \ref{fig_set_16}, we expect to see a
    significant dependence of $F^{ad}$ on the concentrations of the constituents
    in a real system in which the number of electrons in the conduction band can
    be controlled, as in an intermetallic compound for instance.
\end{enumerate}

It is possible to carry out an accurate numerical calculation of the
low--temperature nonadiabatic coefficient $A_{2}$ (see e.g. eq. (25.83) of
\citet{Wallace_72:TOC}). A good estimate can also be obtained from an Einstein
approximation with $\omega_{E}$ determined from $\hbar \omega_{E} = \left< \hbar
\omega \right>$ (\citet{Bock_05}). As shown here (Sec.
\ref{sec_results_phonon_model}), an accurate calculation of $F^{ad}$ requires
the use of realistic phonon frequencies and eigenvectors. While an Einstein
model does give the correct temperature dependence of $F^{ad}$, it does not give
a reliable magnitude (eqs. (28.44) and (28.52) and line 3 of Table 27 of
\citet{Wallace_72:TOC}). Based on a numerical evaluation for Pb to around 100 K,
\citet{Grimvall_76} concluded that $S_{ep}$ is negligible for temperatures above
the mean phonon energy. Our results show that this is not the case however,
since at higher temperatures the adiabatic contribution starts to dominate and
the total entropy rises, shown in Fig. \ref{fig_set_30_b}.
\citet{Allen_Hui_80} argued that $C^{ad}$ can become as large as $C_{el}$ for
certain metals at high temperatures. Our results for nearly--free--electron
metals are not inconsistent with such behavior.

\bibliography{Paper}
\bibliographystyle{apsrev}

\end{document}